\documentclass[aps,pre,twocolumn,showpacs, superscriptaddress]{revtex4}

\usepackage{graphicx}
\begin{document}


\title{Adsorption of hydrophobic polyelectrolytes
 as studied by \emph{in situ} high energy X-Ray reflectivity}

\author{Damien Baigl}
\email[]{baigl@chem.scphys.kyoto-u.ac.jp} \affiliation{Laboratoire
de Physique des Fluides Organis\'{e}s, CNRS UMR 7125, Coll\`{e}ge
de France, Paris, France} \affiliation{Dept of physics, Graduate
school of science, Kyoto University, Kyoto 606-8502, Japan}
\author{Marie-Alice Guedeau-Boudeville}
\affiliation{Laboratoire de Physique des Fluides Organis\'{e}s,
CNRS UMR 7125, Coll\`{e}ge de France, Paris, France}
\author{Raymond Ober}
\affiliation{Laboratoire de Physique des Fluides Organis\'{e}s,
CNRS UMR 7125, Coll\`{e}ge de France, Paris, France}
\author{Fran\c{c}ois Rieutord}
\affiliation{DRFMC-SI3M, CEA-Grenoble and ESRF BM32 beamline,
Grenoble, France}
\author{Michele Sferrazza}
\affiliation{D\'{e}partement de Physique, Universit\'{e} Libre de
Bruxelles, Bruxelles, Belgium}
\author{Olivier Th\'{e}odoly}
\affiliation{Complex Fluids Laboratory, CNRS/Rhodia UMR 166,
Cranbury, NJ 08512, USA}
\author{Thomas A. Waigh}
\affiliation{Department of Physics and Astronomy, University of
Leeds, Leeds, LS2 9JT, UK}
\author{Claudine E. Williams}
\email[]{claudine.williams@college-de-france.fr}
\affiliation{Laboratoire de Physique des Fluides Organis\'{e}s,
CNRS UMR 7125, Coll\`{e}ge de France, Paris, France}

\date{\today}

\begin{abstract}
A series of well-defined hydrophilic and hydrophobic polyelectrolytes of various chain
lengths $N$ and effective charge fractions $f_{eff}$ have been adsorbed onto
oppositely charged solid surfaces immersed in aqueous solutions.
\emph{In situ} high energy X-ray reflectivity has provided the
thickness $h$, the electron density and the roughness of the
adsorbed layer in its aqueous environment. In the case of hydrophobic polyelectrolytes,
we have found $h\propto N^0 f_{eff}^{-2/3}$, in agreement with a pearl-necklace conformation
for the chains induced by a Rayleigh-like instability.
\end{abstract}


\maketitle

\section{Introduction}
Polyelectrolytes are macromolecules containing ionizable groups
which, in a polar solvent like water, dissociate into charges tied
to the polymer backbone and counter-ions dispersed in the solution. They
are called hydrophobic when water is a poor solvent for the
backbone. As amphiphilic water-soluble macromolecules, hydrophobic
polyelectrolytes are of great interest for industrial applications;
 in nature, many biological macromolecules, proteins for
instance, have some intrinsic hydrophobicity. However, even if
experiments~\cite{Baigl_thesis,Lee_2002,Rawiso_GRC2003},
theories~\cite{Kantor,Dobrynin_1996,Dobrynin_1999} and
simulations~\cite{Micka_1999,Limbach_2003} are now all consistent
with a pearl-necklace conformation for the single chain, the
physics of hydrophobic polyelectrolytes is still far from being
fully understood. Indeed, the  role of counter-ions, the
long-range electrostatic interactions, the short-range
monomer-monomer interactions and solvent effects make simulations
and theories difficult to achieve. On the experimental side, the
combination of poor contrast in scattering experiments and
fluctuations of concentration with a wide range of length scales
(from a few nanometers for the pearl size up to a micron for the
Debye length in pure water) make the interpretation of bulk
properties in salt-free solutions
 rather delicate~\cite{Baigl_corlength,Baigl_corlength_Macromol}. Adsorbing chains onto
solid surfaces is an interesting way to freeze the fluctuations of
concentration and eventually those of
conformation~\cite{Limbach_2002}. \emph{In situ} characterization
of the solid-liquid interface can then provide an insight into the
properties of the  chains trapped within the adsorbed layer.
Beside techniques such as AFM~\cite{Kiriy_2002} and
 \emph{in situ} ellipsometry~\cite{Styrkas_2000} which provide the local surface topography and the average adsorbed amount, respectively, X-ray reflectivity
has the advantage of providing the electron density
profile~\cite{Daillant,Plech_2002}. So far, \emph{in situ} X-ray
reflectivity
 has been applied to only a few systems such as lamellar phases~\cite{deJeu_2003} or
polyelectrolyte multilayers~\cite{Plech_2000}.
However, to our knowledge, it has never been used to characterize a polyelectrolyte monolayer in its aqueous environment at the solid-liquid interface.

In this paper we present an experimental study of model
hydrophobic polyelectrolyte monolayers adsorbed onto oppositely
charged or hydrophobic solid surfaces. Adsorbed layers were
characterized \emph{in situ}, \emph{i.e.}, inside the aqueous
solution at the solid-liquid interface, by the technique of X-ray
reflectivity with high energy photons. In a previous study, we
have used \emph{in situ} ellipsometry to investigate the
properties of hydrophobic polyelectrolyte monolayers adsorbed onto
oppositely charged solid surfaces. It has allowed us to establish
the conditions for which the pearl-necklace conformation of the
chains persisted upon adsorption. This required the presence of
added salt in order for the Debye length to be comparable to the
pearl size. In this case, the thickness of the adsorbed layer is
proportional to the pearl size~\cite{Baigl_pearlsize}. For the
present study, the adsorbed layer has been prepared with the same
operating conditions (oppositely charged solid surface and
presence of added salts) and we have measured the thickness of the
adsorbed layer, its roughness and its electron density as a
function of the chemical charge fraction and the length of the
chains. In order to investigate the effect of the solvent quality,
we have made a parallel study of the adsorption of a model
hydrophilic polyelectrolyte. Finally, since hydrophobic
polyelectrolytes are amphiphilic molecules~\cite{Theodoly_EPJE},
we present preliminary
 results for the adsorption onto neutral hydrophobic solid surfaces.

\section{Experimental section}

Throughout this paper, the following notations will be used: $h$
for thickness, $\rho$ for electron density, $\sigma$ for
roughness, Si for silicon, SiO2 for silica, SAM for Self-Assembled
Monolayer, PCS for oppositely charged surface, HS for hydrophobic
surface, PSS for the hydrophobic polyelectrolyte
(poly(styrene-\emph{co}-styrenesulfonate)) and AMAMPS
(poly(acrylamide-\emph{co}-acrylamidomethylpropanesulfonate)) for
the hydrophilic polyelectrolyte.

\subsection{Materials}
\subsubsection{Hydrophobic polyelectrolytes}
\begin{figure}
\includegraphics[width=8.5cm]{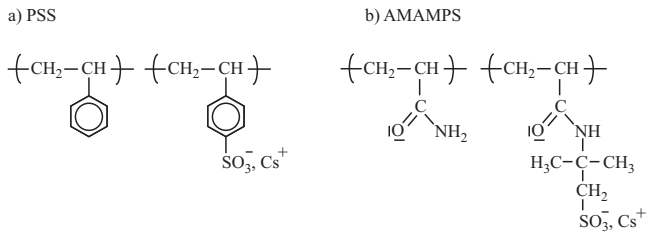}%
\caption{Chemical structures of a) PSS, a model hydrophobic
polyelectrolyte and b) AMAMPS, a model hydrophilic
polyelectrolyte. Both are random copolymers.}\label{PE_molecules}
\end{figure}

As a model hydrophobic polyelectrolyte we have used
poly(styrene-\emph{co}-styrenesulfonate, cesium salt), abbreviated
PSS (see figure~\ref{PE_molecules}a). This random copolymer is
soluble in water when it contains more than 30\% of
styrenesulfonate monomers on average. Therefore, the resulting
macromolecule is a water-soluble charged polymer having a very
hydrophobic backbone of polystyrene and it can be considered as a
model hydrophobic polyelectrolyte. A series of well-defined
monodisperse PSS of various chain lengths $N$ and chemical charge
fractions $f$, \emph{i.e.}, molar percentage of styrenesulfonate
per chain, have been synthesized and characterized according to a
procedure described elsewhere~\cite{Baigl_2002}. Previous osmotic
pressure and freezing point depression measurements have shown a
strong reduction of the effective charge fraction $f_{eff}$ as a
function of $f$. We have found that $f_{eff}$ obeys the following
empirical renormalization law~\cite{Baigl_essafi,Baigl_thesis}:
\begin{equation}\label{eq_feff}
f_{eff}(\%)=100\frac{f-f^*}{100-f^*} \frac{a}{l_B}=\frac{f(\%)-18}{82} 36
\end{equation}
where $f^*$ is the chemical charge fraction (18\%) at which $f_{eff}$ equals 0, $a$ is the monomer size (0.25 nm)
and $l_B$ is the Bjerrum length (0.71 nm in pure water at 25$^\circ$C).
 The characteristics of PSS samples used in this study are summarized
in Table~\ref{PSS_samples}.

\begin{table}
\caption{PSS samples of various chain lengths $N$, chemical charge
fractions $f$ and effective charge fractions $f_{eff}$. $f_{eff}$ is
estimated from $f$ using eq.~\ref{eq_feff}}\label{PSS_samples}
\begin{ruledtabular}
\begin{tabular}{ccc|ccc}
$N$   &$f$(\%)  &$f_{eff}$(\%)  &$N$  &$f$(\%)  &$f_{eff}$(\%)\\
\hline 930 &33 &6.6 &1320     &36 &7.9\\
930 &41 & 10.1  &1320    &53 &15.4\\
930 &46 &12.3   &1320    &71 &23.3\\
930 &62 &19.3   &1320    &91 &32.0\\
\cline{4-6}930 &83 &  28.5 &2520 &37 &8.3\\

    &   &       &2520    &54 &15.8\\
    &   &       &2520    &89 &31.2\\
\end{tabular}
\end{ruledtabular}
\end{table}

\subsubsection{Hydrophilic polyelectrolytes}
As a model hydrophilic polyelectrolyte we have used
poly(acrylamide-\emph{co}-acrylamidomethylpropanesulfonate, cesium
salt), abbreviated AMAMPS (see figure~\ref{PE_molecules}b) because
its polyacrylamide backbone is water soluble. Here the chemical
charge fraction $f$ is the molar percentage of
acrylamidomethylpropanesulfonate monomer per chain and it was
measured by $^1$H NMR (Bruker Avance300 spectrometer) in deuterium
oxide (Aldrich). AMAMPS's with $f$ ranging from 34\% to 100\% have
been synthesized by radical copolymerization of acrylamide
(Aldrich) and sodium 2-acrylamido-2-methyl-1-propanesulfonate,
initiated by potassium persulfate and tetradimethylethylenediamine
(Aldrich), in a 60/20 water-ethanol mixture according to a
procedure inspired from~\cite{McCormck_1982} and fully described
in~\cite{Baigl_thesis}.

\subsubsection{Surfaces}
\begin{figure}
\includegraphics[width=8.5cm]{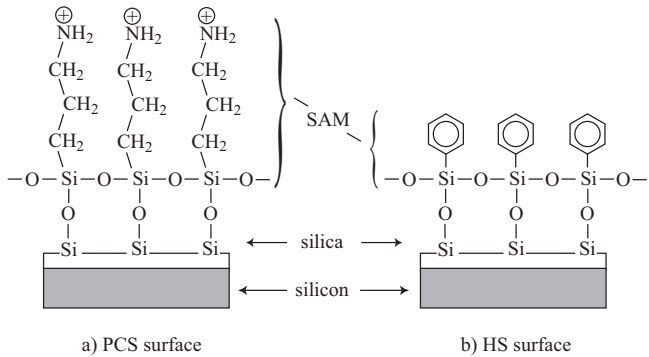}
\caption{Chemical structures of the planar solid surfaces. a)
Positively charged surface (PCS) and b) hydrophobic surface
(HS).}\label{surfaces}
\end{figure}

Two types of solid surfaces have been prepared, as pictured in
figure~\ref{surfaces}: a) positively charged surface (PCS) and b)
neutral hydrophobic surface (HS). For this purpose, self-assembled
monolayers (SAM) have been grafted on the native silica layers
covering silicon wafers (Siltronix, Archamps, France; diameter:
25.4 mm, thickness: 2 mm). First, each wafer is cleaned by a 45
minutes UV-O$_3$ treatment prior to 15 minutes exposure under an
oxygen flow saturated with water. Right after this activation
process, the wafer is immersed in 15 mL of a freshly prepared
silane (Aminopropyltrimethoxysilane (a) or Phenyltrimethoxysilane
(b), Aldrich) solution at 0.15 mol.L$^{-1}$ in anhydrous toluene
(stored on molecular sieves) and left under gentle stirring during
15 minutes at ambient temperature. The wafer is then cautiously
withdrawn to be rinsed thoroughly by pure anhydrous toluene
followed by pure anhydrous chloroform. After rinsing, the wafer is
dried with gaseous nitrogen (Air Liquide), left in a vacuum oven
at 90$^\circ$C for 12 minutes, rinsed again by pure anhydrous
toluene and chloroform and dried under nitrogen flow. The wafer is
finally stored under nitrogen in an airtight box. The silica layer
thickness $h_{SiO2}$ was measured by ellipsometry in air after
activation and before silanation. The SAM thickness $h_{SAM}$ was
measured by ellipsometry in air while its roughness $\sigma_{SAM}$
and density $\rho_{SAM}$ were measured by X-ray reflectivity in
air at our laboratory. The topography was checked by atomic force
microscopy. All these measurements are consistent with a dense
self-assembled monolayer ($h_{SAM}=1.1~$nm for PCS and 0.4 nm for
HS, $\rho_{SAM}=0.29~$ e$^-\mathrm{\AA}^{-3}$,
$\sigma_{SAM}=0.5~$nm).

The adsorbed polyelectrolyte layers were prepared by spontaneous adsorption from aqueous solution
of polymer at 0.01 mol.L$^{-1}$ and CsCl at 0.1 mol.L$^{-1}$. In these conditions it has been
shown that the pearl-necklace conformation persists in the adsorbed state~\cite{Baigl_pearlsize}.

\subsection{\emph{in situ} X-ray reflectivity}
\subsubsection{Principles}
The specular reflectivity $R$ is defined as the ratio of the
reflected intensity to the intensity of the incident beam. An
X-ray reflectivity experiment consists in measuring $R$ as a
function of the incident angle $\theta$ or of the vertical
momentum transfer $q$ ($q=4\pi n/\lambda\sin{\theta}$ where
$\lambda$ is the wavelength). The refractive index $n$ of matter
for X rays is given by $n=1-\delta-i\beta$ where
$\delta\approx\rho r_e\lambda^2/(2\pi)$, $\rho$ is the electron
density and $r_e$ the classical electron radius, and
$\beta=\mu\lambda/4\pi$ is related to the absorption coefficient
$\mu$. For X-ray wavelengths, $\delta$ and $\beta$ are much less
than 1. Neglecting absorption, an X-ray wave propagating in a
medium 0 is totally reflected on a substrate 1 for
$\theta<\theta_c$ ($q<q_c$ respectively), where the critical angle
$\theta_c$ is given by
\begin{equation}
\theta_c\approx \sqrt{2(\delta_{1}-\delta_{0})}
\end{equation}
For $\theta\geq \theta_c$, in the case of a sharp ideal interface, the reflectivity, called Fresnel reflectivity $R_F$, is approximately
\begin{equation}\label{eq_qmoins4}
R_F\approx \frac{\theta_c^4}{16\theta^4}= \frac{q_c^4}{16q^4}
\end{equation}
For real surfaces, the reflectivity can be expressed as a function of the electron density profile~\cite{Pershan_1984}:
\begin{equation}
R=R_F\left|\int\frac{1}{\rho_0}\frac{d\rho}{dz}\exp(-iqz)dz\right|^2
\end{equation}

\subsubsection{Reflectivity experiment}
Experimental conditions have been optimized
regarding the strong absorption of X-ray radiation by water. The absorption coefficient $\alpha$, defined as the ratio
 of the transmitted intensity $I_t$ after travelling a distance $d$ to incident intensity $I_0$,
can be expressed as $\alpha=I_t/I_0=\exp(-\mu d)$. Since $\mu$ is
a decreasing function of energy, high energy synchrotron radiation
is necessary~\cite{note_absorption}. However, working angles
decrease with energy since $\theta_c\propto E^{-1}$.
 Therefore, the longitudinal length $L$ of the corresponding footprint at $\theta_c$
(in the direction of beam propagation), $L=h/\sin{\theta_c}$, is an increasing function of $E$. Larger wafers are thus required at higher energy, implying an
 increase of $d$ and $\alpha$. The final choice of our experimental conditions was a compromise between these requirements as described below.

Experiments were performed on BM32 beamline of the European
Synchrotron Radiation Facility (ESRF) in Grenoble (France) at a photon energy of 27 keV
($\lambda\approx 0.046~$nm) corresponding to a critical angle $\theta_c \approx 0.048^\circ$ ($q_c\approx 0.23~$nm$^{-1}$)~\cite{note_energie}. A point, low background scintillation
detector has been used. The high precision beamline goniometer has been used in the range $0\leq \theta \leq 0.76^\circ$ ($0\leq q\leq 3.6$~nm$^{-1}$). Working at such small angles requires perfectly planar surfaces. Hence, before each \emph{in situ} reflectivity experiment,
 the planeity of the solid surface has been checked by measuring the width of
 the reflected beam. The incident beam has been limited by slits to a size of  $l=1~$mm (horizontal width) by $h=20~\mu$m (vertical height).
 The corresponding footprint at $\theta_c$ had
  a longitudinal length $L\approx 23.9$~mm, which is
slightly smaller than the wafer size (25.4~mm).
\begin{figure}
\includegraphics[width=8.5cm]{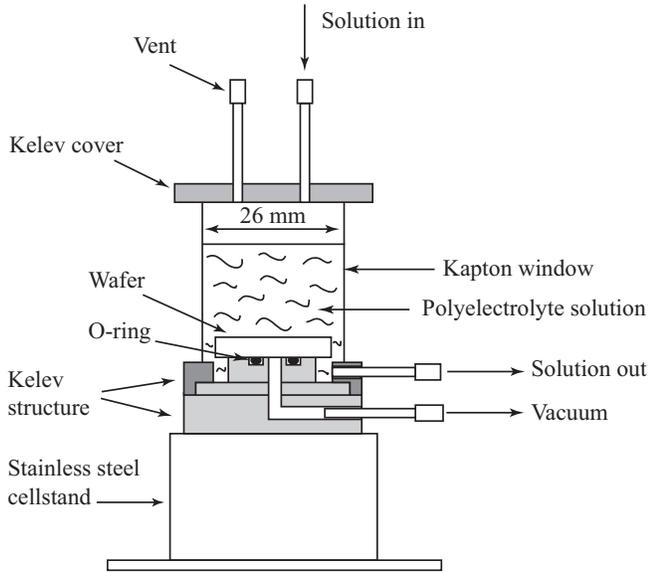}
\caption{Liquid cell for the reflectivity experiment.}\label{Schema_cellule}
\end{figure}
The liquid cell, especially designed for these conditions, is illustrated in figure~\ref{Schema_cellule}. Its diameter, 26 mm, is just above that of the solid surface.
 The watertight cylindrical structure is made of Kelev. This hydrophobic polymer material prevents chemical and ionic contamination.
 Windows have been made of Kapton, a non-absorbing material with negligible scattering. Since the solid surface is maintained by depression, it is important to use thick
wafers (2 mm) to avoid deformation. The in/out vent system allows one to flush the cell with the surface kept immersed.
Finally, with these conditions, the absorption coefficient was approximately $\alpha\approx0.3$.

The background for the reflectivity curve has been measured at an angle of $2\theta + 0.1^\circ$
and substracted afterwards. It was checked that during the typical exposure time of about 15 minutes,
no beam damage of the surface occurred.

\section{Results and discussion}

\subsection{Experimental data}

\begin{figure}
 \includegraphics[width=8.5cm]{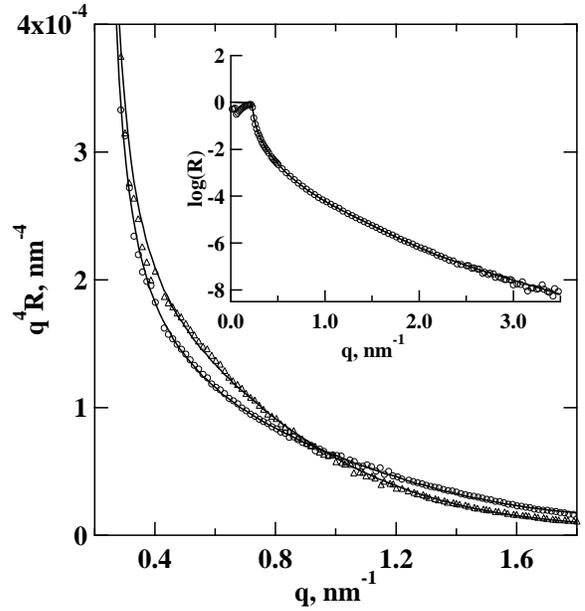}%
 \caption{$q^4R$ as a function of the vertical momentum transfer
$q$ before (open triangles) and after the adsorption (open circles)
of one PSS ($N=2520$, $f=37\%$). Inset: logarithm of reflectivity log(R) as a function of
$q$ for the adsorbed PSS layer ($N=2520$, $f=37\%$). Solid lines
result from fits using Parrat's algorithm (eq.~\ref{eq_kj}-\ref{eq_Rcalc}).}\label{q4R_log_fit}
 \end{figure}

A typical reflectivity scan for the model hydrophobic
polyelectrolyte (PSS) is presented in the inset of
figure~\ref{q4R_log_fit}. The logarithm of reflectivity
log(\emph{R}) is plotted as a function of the vertical momentum
transfer $q$ for an adsorbed PSS layer ($N=2520$, $f=37\%$)
immersed in water. The critical edge at 0.23 nm$^{-1}$ is clearly
visible. For $q<q_c$, $R$ is less than 1 for geometric reasons.
For $q>q_c$, $R$ decreases steeply as a function of $q$. According
to eq.~\ref{eq_qmoins4}, at a  sharp silicon-water ideal
interface, $R$ would be proportional to $q^{-4}$. In the main
graph, $q^4R$ is plotted as a function of $q$ for a PCS surface
before (open triangles) and after (open circles) the adsorption of
the PSS layer ($N=2520$, $f=37\%$). In this representation, the
presence of the PSS layer is clearly evidenced, the $q^4R$ curve
being first below, crossing at $q\approx~1~\mathrm{nm}^{-1}$ and
staying finally above that of the bare surface. Therefore, the
contribution of the PSS layer is qualitatively given by the
difference $\Delta q^4R$ defined as follows~:
\begin{equation}
\Delta
q^4R=q^4R_{\mathrm{after~adsorption}}-q^4R_{\mathrm{before~adsorption}}
\end{equation}

\begin{figure}
\includegraphics[width=8.5cm]{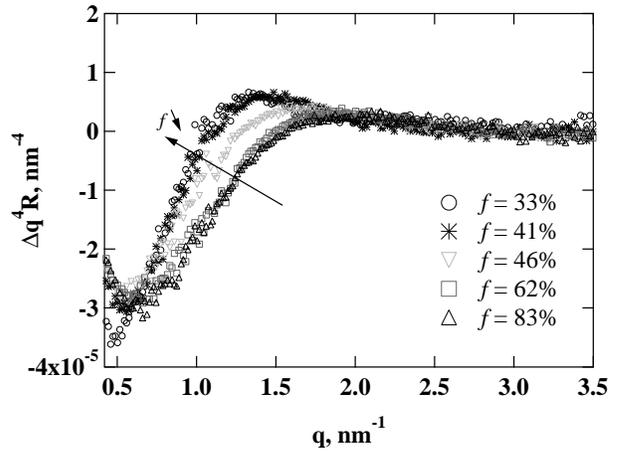}%
\caption{$\Delta q^4R$ as a function of the vertical
momentum transfer $q$ for PSS of chain length
$N=930$ and various chemical charge fractions $f$. The arrow indicates decreasing values of $f$}\label{Deltaq4I}
\end{figure}

\begin{figure}
\includegraphics[width=8.5cm]{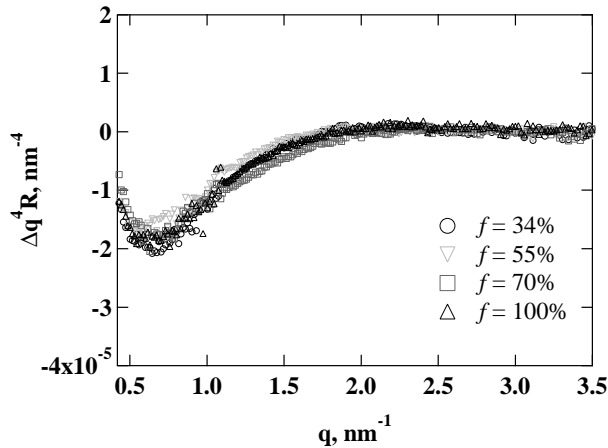}%
\caption{$\Delta q^4R$ as a function of the vertical
momentum transfer $q$ for AMAMPS (hydrophilic polyelectrolyte) of various
chemical charge fractions $f$.}\label{Deltaq4I_AMAMPS}
\end{figure}

$\Delta q^4R$ is plotted in figure~\ref{Deltaq4I} as a function of
$q$ for a series of PSS of a chain length $N=930$ and various
chemical charge fractions $f$. It shows a strong dependence of $\Delta q^4R$, \emph{i. e.},
of the layer characteristics, on $f$.
  For the sake of comparison, the same experiment
has been performed with the model hydrophilic polyelectrolyte called
AMAMPS. In figure~\ref{Deltaq4I_AMAMPS}, $\Delta q^4 R$ is plotted
as a function of $q$ for AMAMPS of various chemical charge
fractions in the same range as those in figure~\ref{Deltaq4I}.
Here the $\Delta q^4R$ curves are superposable, indicating that the properties
of the  adsorbed AMAMPS layer are independent of $f$~\cite{note_AMAMPS}. These experiments show
a clear difference in the adsorption behavior of the hydrophilic polyelectrolytes and the hydrophobic ones.
At this point, one should recall that the bulk solution properties of the same polymers are also markedly different.
For AMAMPS, it has been shown that the structure of the solution is independent of $f$ in the range where
the effective charge fraction is renormalized by Manning condensation to a constant value of 36\%~\cite{Baigl_essafi,Baigl_thesis}.
In contrast, for PSS solutions, the structural characteristics all depend on $f$
and corroborate the pearl-necklace conformation of the single chain (\cite{Baigl_corlength} and references therein). At the same time
the effective charge has been found to be a strong function of $f$ (see eq.~\ref{eq_feff}).
It is tempting to assume that the chain conformation at the surface is related to that in the bulk and that the evolution
 of the adsorbed PSS layer properties on $f$ observed in figure~\ref{Deltaq4I} are related to conformational
effects, \emph{i. e.}, the presence of pearls in the specific case of hydrophobic polyelectrolytes.

In what follows, we will focus on the evolution of the
PSS layer thickness as a function of the effective charge fraction.

\subsection{Data analysis}
Quantitative analysis have been performed by fitting the
experimental reflectivity curves with the recursive Parratt
algorithm~\cite{Parat_1954}. Our system has been modeled as a
stratified medium composed of layers numbered from 0 (water) to
$n$ (silicon). In layer $j$, the propagating vector projection
along the vertical direction is
\begin{equation}\label{eq_kj}
k_j\approx \frac{2\pi}{\lambda}\sqrt{\theta^2+2(\delta_0-\delta_j)+2i(\beta_0-\beta_j)}
\end{equation}

 The amplitude of the reflection coefficient for the electric field $r_{j-1,j}$ at the interface separating
 layer $j-1$ and layer $j$ is given by
\begin{equation}
r_{j-1,j}=a_{j-1}^2\frac{r_{j,j+1}+F_{j-1,j}}{r_{j,j+1}F_{j-1,j}+1}
\end{equation}
where
\begin{equation}
a_{j-1}= \exp(-ik_jh_j)
\end{equation} is the delay produced by the layer $j$ of thickness $h_j$ and $F_{j-1,j}$ the Fresnel coefficient given by
\begin{equation}
F_{j-1,j}=\frac{k_{j-1}-k_j}{k_{j-1}+k_j}\exp(-2k_{j-1}k_j\sigma_{j-1,j}^2)
\end{equation}
introducing the Nevot and Croce roughness $\sigma_{j-1,j}$ of the
interface between layer $j-1$ and layer $j$~\cite{Nevot_1980}.
This
 recursive system is solved knowing $r_{n,n+1}=0$ and the reflectivity is given by
 \begin{equation}\label{eq_Rcalc}
R=|r_{0,1}r_{0,1}^*|^2
 \end{equation}
 where $r_{0,1}^*$ is the complex conjugate of $r_{0,1}$.
The quality of the fit of the experimental data is illustrated by
the solid lines in figure~\ref{q4R_log_fit}. For all the data, the
above algorithm has been performed using three fitting parameters
: the PSS layer thickness $h_{PSS}$, its roughness $\sigma_{PSS}$
and its electron density $\rho_{PSS}$. For each substrate, all
other parameters have been fixed
($\rho_{Si}=0.70~$e$^-\mathrm{\AA}^{-3}$, $\sigma_{Si}=0.5~$nm,
$\rho_{SiO2}=0.67~$e$^-\mathrm{\AA}^{-3}$,
$\rho_{H2O}=0.33~$e$^-\mathrm{\AA}^{-3}$) or measured
independently by ellipsometry ($h_{Si02}$) and \emph{in situ} high
energy reflectivity prior to polyelectrolyte adsorption
($h_{SAM}$, $\rho_{SAM}$, $\sigma_{SAM}$). It is important to note
that, whereas the thickness and roughness of the SAM inside water
has been found to be the same as that measured in air (by
ellipsometry and X-ray reflectivity), its electron density
systematically shifted from 0.29 e$^-\mathrm{\AA}^{-3}$ in air to
0.45 e$^-\mathrm{\AA}^{-3}$ in pure water~\cite{note_density}.
Only this later value allowed to fit the reflectivity profiles of
the PSS layers and was used throughout.

\subsection{Electron density and roughness}
The electron density $\rho_{PSS}$ was found to increase as a function of $f$
from 0.36 to 0.40~e$^-\mathrm{\AA}^{-3}$. This is
to be expected since most of the contrast comes from the counter-ions
in the layer. Indeed, according to the $f_{eff}$ values (listed in table~\ref{PSS_samples}), the amount of condensed counter-ions per chain varies within
a factor of 2 in the $f$ range
explored. Assuming that the adsorbed amount remains constant as verified by \emph{in situ} ellipsometry~\cite{Baigl_pearlsize},
the electron density should increase accordingly.  On the other hand, the roughness of the PSS layer has been measured to be between 1.0 and 1.5 nm.
 To be meaningful, this value has to be compared to the thickness of the adsorbed layer, typically between 1 and 5 nm (see next subsection).
 There is thus a strong coupling between the roughness and the thickness and a precise determination of each parameter independently is difficult.
 Nevertheless, there is a clear evidence of the rough character of the adsorbed layer at a molecular level. This is qualitatively in agreement
 with the presence of pearls composing the adsorbed PSS layer.

\subsection{Thickness}

\begin{figure}
\includegraphics[width=8.5cm]{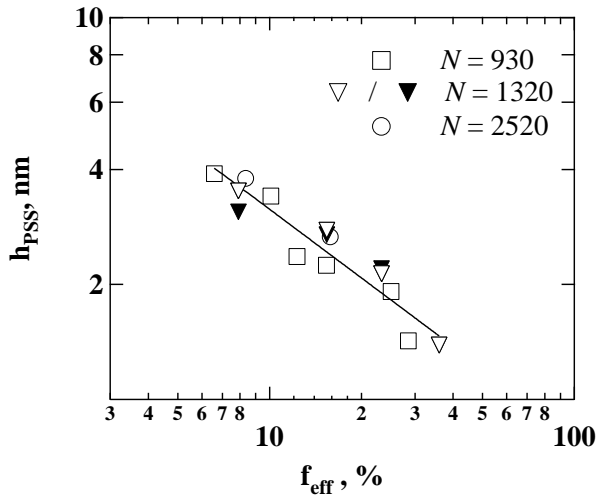}%
\caption{Thickness $h_{PSS}$ of the adsorbed PSS layer as a
function of the effective charge fraction $f_{eff}$ for various
chain lengths $N$. The straight line has a slope of -2/3. The open
symbols correspond to the adsorption onto oppositely charge
surfaces, the filled inverted triangles to the adsorption onto
hydrophobic surfaces. Errors on experimental points are around
0.4~nm.}\label{Thickness}
\end{figure}

In figure~\ref{Thickness}, the thickness $h_{PSS}$ of the PSS layers, is plotted
as a function of the effective charge fraction $f_{eff}$. Let us first consider the adsorption of PSS onto oppositely
charged surfaces (open symbols).
$h_{PSS}$ decreases with $f_{eff}$ and is independent of $N$ at least
for the large molecules considered here ($N \geq 930$)~\cite{note_chainlength}. Since $\rho_{PSS}$ and
$\sigma_{PSS}$ appear only slightly dependent on $f$, the strong evolution with $f$ observed in figure~\ref{Deltaq4I} is mostly due
to variations of $h_{PSS}$. As shown in figure~\ref{Thickness}, $h_{PSS}$ and $f_{eff}$ span slightly less than
a decade. It is thus impossible to determine whether a power law exists and what the precise value of the $f_{eff}$ exponent would be. Nevertheless, within
experimental accuracy, the $h_{PSS}$ dependence on $f_{eff}$ is in agreement with the predicted power law:
\begin{equation}\label{eq_hPSS}
h_{PSS}\propto a N^0 f_{eff}^{-2/3}
\end{equation}
as represented by the straight line in figure~\ref{Thickness}. For all films, the thickness was the same in the adsorbing solution
or after flushing with pure water, indicating that all adsorbed chains are strongly attached to the surface and do not desorb. On the other hand,
the results of figure~\ref{Thickness} and eq.~\ref{eq_hPSS} are in perfect agreements with previous results from  \emph{in situ} ellipsometry experiments performed
on the same system~\cite{Baigl_pearlsize}.
In contrast to ellipsometry, X-ray reflectivity does not require any assumption on the refractive index and allows one
to establish the electron density profile. X-ray reflectivity is thus a reliable and accurate technique to
fully characterize a polyelectrolyte monolayer at a solid-liquid interface.

It is worth recalling that we have chosen the conditions of adsorption in order to estimate the pearl size. For this purpose,
the adsorption has been made in the presence of added salts so that the Debye length is comparable to the pearl size.
As it has been shown experimentally~\cite{Baigl_pearlsize} and  predicted theoretically~\cite{Dobrynin_2002}, the pearl-necklace
 conformation persists upon adsorption in these conditions of screened electrostatic attraction to the surface. Furthermore,
 the electrostatic repulsion between two neighbouring pearls is sufficiently screened to induce a compaction of the
 pearl-necklace on its pearls~\cite{note_perles_sel}. Therefore, the PSS layer can be viewed as dense carpet of pearls. The thickness $h_{PSS}$
 is closely related to the pearl size $D_p$ and we assume that $h_{PSS}$ is proportional $D_p$:
 \begin{equation}\label{eq_hPSSDp}
h_{PSS} \propto D_p
 \end{equation}
By analogy with the Rayleigh
 instability of a charged droplet~\cite{Rayleigh_1882}, the pearl-size is predicted~\cite{Dobrynin_1996} to
 scale as:
 \begin{equation}\label{eq_Dp}
D_p \propto a N^0 \lambda^{-2/3}
 \end{equation}
where $\lambda$ is the linear charge density along the chain. For a real polyelectrolyte system, we assume
that $\lambda$ is given by the effective charge fraction $f_{eff}$ and we expect:
\begin{equation}\label{eq_Dpfeff}
D_p \propto a N^0 f_{eff}^{-2/3}
\end{equation}
Therefore, the experimental results of figure~\ref{Thickness} and
eq.~\ref{eq_hPSS} are in perfect agreement with the scaling
predictions of the pearl-necklace model for the pearl size
(eq.~\ref{eq_Dpfeff}). This is an indirect experimental evidence
of the pearl-necklace conformation induced by a Rayleigh-like
instability. This confirms that the effective
 charge fraction $f_{eff}$, rather than $f$, is also controlling the intra-chain electrostatic interactions.

\subsection{Adsorption onto hydrophobic surfaces }

All previous results concerned the adsorption on positively charged
surfaces (PCS). We have to be concerned with the deformation of pearls upon adsorption since a strong flattening
of pearls has also been predicted in the case of unscreened attraction~\cite{Borisov_2001}.  For this purpose, we have also studied, in a less extended way,
 the adsorption of PSS onto hydrophobic surfaces (HS). This
corresponds to the filled symbols in figure~\ref{Thickness}.
Before all, it is important to note that the kinetics of
adsorption differs largely from that of electrostatic
adsorption~\cite{Theo_thesis}. In the former case, chains have to
diffuse towards the HS and the adsorption process is slow whereas
in the latter, the adsorption to the PCS is driven by
electrostatic attraction. In order to measure the equilibrium
thickness in the case of HS, $h_{PSS}$ has been measured after a
contact time of 24 hours between the surface and the PSS solution.
Data of figure~\ref{Thickness} are preliminary results and only
three points have been obtained. Nevertheless, within experimental
accuracy, it seems that the behaviour on HS surface is identical
to that on PCS. In other words, the pearl-necklace conformation
seems to persist also upon adsorption onto hydrophobic surfaces.
We are now pursuing this study and we expect that further
experiments will elucidate this interesting point.

\section{Conclusion}
The model hydrophobic polyelectrolyte poly(styrene-\emph{co}-styrene sulfonate), called PSS, of various chain lengths $N$ and effective charge fractions $f_{eff}$
has been adsorbed onto oppositely charged surfaces immersed in water. An original technique, \emph{in situ} high energy X-ray reflectivity,
 has allowed us to measure the electron density $\rho_{PSS}$, the roughness $\sigma_{PSS}$ and the thickness $h_{PSS}$ of the PSS
 monolayer at the buried solid-liquid interface.
  In the presence of an adequate amount of added salts, all measured parameters are a strong function of the charge fraction in marked contrast
  to the case of a hydrophilic (AMAMPS) adsorbed in the same conditions. We have found
$h_{PSS}\propto aN^0f_{eff}^{-2/3}$ in agreement with the scaling prediction for the pearl-size $D_p$
in the pearl-necklace model if one interprets $h_{app}$ as a measure of $D_p$. Preliminary investigations of PSS layers adsorbed
on a hydrophobic solid surface at the same ionic strength point to a very similar structure of the layers. Further experiments will analyze
the off-specular reflectivity in order to characterize the in-plane structure of the PSS layer.

\begin{acknowledgments}
It is a pleasure to thank Jean Daillant for illuminating
discussions about the intricacies of X-ray reflectivity and for
his active help when first setting up the experiment. We also wish
to thank Victor Etgens for sharing with us his ideas about the
design of the cell. The efficient help of the technical and
scientific staff of ESRF, particulary O. Konovalov, B. Jean and O.
Plantevin, is gratefully acknowledge. C. Monteux kindly assisted
us for one set of experimental runs.
\end{acknowledgments}

\bibliography{BibrefX}

\begin{thebibliography}{37}
\expandafter\ifx\csname natexlab\endcsname\relax\def\natexlab#1{#1}\fi
\expandafter\ifx\csname bibnamefont\endcsname\relax
  \def\bibnamefont#1{#1}\fi
\expandafter\ifx\csname bibfnamefont\endcsname\relax
  \def\bibfnamefont#1{#1}\fi
\expandafter\ifx\csname citenamefont\endcsname\relax
  \def\citenamefont#1{#1}\fi
\expandafter\ifx\csname url\endcsname\relax
  \def\url#1{\texttt{#1}}\fi
\expandafter\ifx\csname urlprefix\endcsname\relax\def\urlprefix{URL }\fi
\providecommand{\bibinfo}[2]{#2}
\providecommand{\eprint}[2][]{\url{#2}}

\bibitem[{\citenamefont{Baigl}(2003)}]{Baigl_thesis}
\bibinfo{author}{\bibfnamefont{D.}~\bibnamefont{Baigl}}, Ph.D. thesis,
  \bibinfo{school}{Paris VI}, \bibinfo{address}{Paris, France}
  (\bibinfo{year}{2003}),
  \bibinfo{note}{http://tel.ccsd.cnrs.fr/documents/archives0/00/
  00/36/20/tel-00003620-00/tel-00003620.pdf}.

\bibitem[{\citenamefont{Lee et~al.}(2002)\citenamefont{Lee, Green, Mike\u{s},
  and Morawetz}}]{Lee_2002}
\bibinfo{author}{\bibfnamefont{M.-J.} \bibnamefont{Lee}},
  \bibinfo{author}{\bibfnamefont{M.~M.} \bibnamefont{Green}},
  \bibinfo{author}{\bibfnamefont{F.}~\bibnamefont{Mike\u{s}}},
  \bibnamefont{and} \bibinfo{author}{\bibfnamefont{H.}~\bibnamefont{Morawetz}},
  \bibinfo{journal}{Macromolecules} \textbf{\bibinfo{volume}{35}},
  \bibinfo{pages}{4216} (\bibinfo{year}{2002}).

\bibitem[{\citenamefont{Rawiso}()}]{Rawiso_GRC2003}
\bibinfo{author}{\bibfnamefont{M.}~\bibnamefont{Rawiso}},
  \bibinfo{note}{private communication, July 2003}.

\bibitem[{\citenamefont{Kantor and Kardar}(1994)}]{Kantor}
\bibinfo{author}{\bibfnamefont{Y.}~\bibnamefont{Kantor}} \bibnamefont{and}
  \bibinfo{author}{\bibfnamefont{M.}~\bibnamefont{Kardar}},
  \bibinfo{journal}{Europhys. Lett.} \textbf{\bibinfo{volume}{27}},
  \bibinfo{pages}{643} (\bibinfo{year}{1994}).

\bibitem[{\citenamefont{Dobrynin et~al.}(1996)\citenamefont{Dobrynin,
  Rubinstein, and Obukhov}}]{Dobrynin_1996}
\bibinfo{author}{\bibfnamefont{A.~V.} \bibnamefont{Dobrynin}},
  \bibinfo{author}{\bibfnamefont{M.}~\bibnamefont{Rubinstein}},
  \bibnamefont{and} \bibinfo{author}{\bibfnamefont{S.~P.}
  \bibnamefont{Obukhov}}, \bibinfo{journal}{Macromolecules}
  \textbf{\bibinfo{volume}{29}}, \bibinfo{pages}{2974} (\bibinfo{year}{1996}).

\bibitem[{\citenamefont{Dobrynin and Rubinstein}(1999)}]{Dobrynin_1999}
\bibinfo{author}{\bibfnamefont{A.~V.} \bibnamefont{Dobrynin}} \bibnamefont{and}
  \bibinfo{author}{\bibfnamefont{M.}~\bibnamefont{Rubinstein}},
  \bibinfo{journal}{Macromolecules} \textbf{\bibinfo{volume}{32}},
  \bibinfo{pages}{915} (\bibinfo{year}{1999}).

\bibitem[{\citenamefont{Micka et~al.}(1999)\citenamefont{Micka, Holm, and
  Kremer}}]{Micka_1999}
\bibinfo{author}{\bibfnamefont{U.}~\bibnamefont{Micka}},
  \bibinfo{author}{\bibfnamefont{C.}~\bibnamefont{Holm}}, \bibnamefont{and}
  \bibinfo{author}{\bibfnamefont{K.}~\bibnamefont{Kremer}},
  \bibinfo{journal}{Langmuir} \textbf{\bibinfo{volume}{15}},
  \bibinfo{pages}{4033} (\bibinfo{year}{1999}).

\bibitem[{\citenamefont{Limbach and Holm}(2003)}]{Limbach_2003}
\bibinfo{author}{\bibfnamefont{H.-J.} \bibnamefont{Limbach}} \bibnamefont{and}
  \bibinfo{author}{\bibfnamefont{C.}~\bibnamefont{Holm}}, \bibinfo{journal}{J.
  Phys. Chem. B} \textbf{\bibinfo{volume}{107}}, \bibinfo{pages}{8041}
  (\bibinfo{year}{2003}).

\bibitem[{\citenamefont{Baigl et~al.}(2003{\natexlab{a}})\citenamefont{Baigl,
  Ober, Qu, Fery, and Williams}}]{Baigl_corlength}
\bibinfo{author}{\bibfnamefont{D.}~\bibnamefont{Baigl}},
  \bibinfo{author}{\bibfnamefont{R.}~\bibnamefont{Ober}},
  \bibinfo{author}{\bibfnamefont{D.}~\bibnamefont{Qu}},
  \bibinfo{author}{\bibfnamefont{A.}~\bibnamefont{Fery}}, \bibnamefont{and}
  \bibinfo{author}{\bibfnamefont{C.~E.} \bibnamefont{Williams}},
  \bibinfo{journal}{Europhys. Lett.} \textbf{\bibinfo{volume}{62}},
  \bibinfo{pages}{588} (\bibinfo{year}{2003}{\natexlab{a}}).

\bibitem[{\citenamefont{Qu et~al.}(2003)\citenamefont{Qu, Baigl, Williams,
  Mohwald, and Fery}}]{Baigl_corlength_Macromol}
\bibinfo{author}{\bibfnamefont{D.}~\bibnamefont{Qu}},
  \bibinfo{author}{\bibfnamefont{D.}~\bibnamefont{Baigl}},
  \bibinfo{author}{\bibfnamefont{C.~E.} \bibnamefont{Williams}},
  \bibinfo{author}{\bibfnamefont{H.}~\bibnamefont{Mohwald}}, \bibnamefont{and}
  \bibinfo{author}{\bibfnamefont{A.}~\bibnamefont{Fery}},
  \bibinfo{journal}{Macromolecules} \textbf{\bibinfo{volume}{36}},
  \bibinfo{pages}{6878} (\bibinfo{year}{2003}).

\bibitem[{\citenamefont{Limbach et~al.}(2002)\citenamefont{Limbach, Holm, and
  Kremer}}]{Limbach_2002}
\bibinfo{author}{\bibfnamefont{H.-J.} \bibnamefont{Limbach}},
  \bibinfo{author}{\bibfnamefont{C.}~\bibnamefont{Holm}}, \bibnamefont{and}
  \bibinfo{author}{\bibfnamefont{K.}~\bibnamefont{Kremer}},
  \bibinfo{journal}{Europhys. Lett.} \textbf{\bibinfo{volume}{60}},
  \bibinfo{pages}{566} (\bibinfo{year}{2002}).

\bibitem[{\citenamefont{Kyriy et~al.}(2002)\citenamefont{Kyriy, Gorodyska,
  Minko, Jaeger, \u{S}t\u{e}p\`{a}nek, and Stamm}}]{Kiriy_2002}
\bibinfo{author}{\bibfnamefont{A.}~\bibnamefont{Kyriy}},
  \bibinfo{author}{\bibfnamefont{G.}~\bibnamefont{Gorodyska}},
  \bibinfo{author}{\bibfnamefont{S.}~\bibnamefont{Minko}},
  \bibinfo{author}{\bibfnamefont{W.}~\bibnamefont{Jaeger}},
  \bibinfo{author}{\bibfnamefont{P.}~\bibnamefont{\u{S}t\u{e}p\`{a}nek}},
  \bibnamefont{and} \bibinfo{author}{\bibfnamefont{M.}~\bibnamefont{Stamm}},
  \bibinfo{journal}{J. Am. Chem. Soc.} \textbf{\bibinfo{volume}{124}},
  \bibinfo{pages}{13454} (\bibinfo{year}{2002}).

\bibitem[{\citenamefont{Styrkas et~al.}(2000)\citenamefont{Styrkas, Lu, Keddie,
  and Armes}}]{Styrkas_2000}
\bibinfo{author}{\bibfnamefont{B.~V.} \bibnamefont{Styrkas},
  \bibfnamefont{D.~A.}}, \bibinfo{author}{\bibfnamefont{J.~R.}
  \bibnamefont{Lu}}, \bibinfo{author}{\bibfnamefont{J.~L.}
  \bibnamefont{Keddie}}, \bibnamefont{and}
  \bibinfo{author}{\bibfnamefont{S.~P.} \bibnamefont{Armes}},
  \bibinfo{journal}{Langmuir} \textbf{\bibinfo{volume}{16}},
  \bibinfo{pages}{5980} (\bibinfo{year}{2000}).

\bibitem[{\citenamefont{Daillant and Gibaud}(1999)}]{Daillant}
\bibinfo{editor}{\bibfnamefont{J.}~\bibnamefont{Daillant}} \bibnamefont{and}
  \bibinfo{editor}{\bibfnamefont{A.}~\bibnamefont{Gibaud}}, eds.,
  \emph{\bibinfo{title}{X-ray and neutron reflectivity: principles and
  applications}} (\bibinfo{publisher}{Springer, Berlin}, \bibinfo{year}{1999}).

\bibitem[{\citenamefont{Plech and Salditt}(2002)}]{Plech_2002}
\bibinfo{author}{\bibfnamefont{A.}~\bibnamefont{Plech}} \bibnamefont{and}
  \bibinfo{author}{\bibfnamefont{T.}~\bibnamefont{Salditt}}, in
  \emph{\bibinfo{booktitle}{Handbook of polyelectrolytes and their
  applications}} (\bibinfo{address}{Stevenson Ranch, CA, USA},
  \bibinfo{year}{2002}), vol.~\bibinfo{volume}{1}, p. \bibinfo{pages}{265}.

\bibitem[{\citenamefont{de~Jeu}(2003)}]{deJeu_2003}
\bibinfo{author}{\bibfnamefont{W.~H.} \bibnamefont{de~Jeu}},
  \bibinfo{journal}{Rev. Mod. Phys.} \textbf{\bibinfo{volume}{75}},
  \bibinfo{pages}{181} (\bibinfo{year}{2003}).

\bibitem[{\citenamefont{Plech et~al.}(2000)\citenamefont{Plech, Salditt,
  M\"{u}nster, and Peisl}}]{Plech_2000}
\bibinfo{author}{\bibfnamefont{A.}~\bibnamefont{Plech}},
  \bibinfo{author}{\bibfnamefont{T.}~\bibnamefont{Salditt}},
  \bibinfo{author}{\bibfnamefont{C.}~\bibnamefont{M\"{u}nster}},
  \bibnamefont{and} \bibinfo{author}{\bibfnamefont{J.}~\bibnamefont{Peisl}},
  \bibinfo{journal}{J. Colloid Interface Sci.} \textbf{\bibinfo{volume}{223}},
  \bibinfo{pages}{74} (\bibinfo{year}{2000}).

\bibitem[{\citenamefont{Baigl et~al.}(2003{\natexlab{b}})\citenamefont{Baigl,
  Sferrazza, and Williams}}]{Baigl_pearlsize}
\bibinfo{author}{\bibfnamefont{D.}~\bibnamefont{Baigl}},
  \bibinfo{author}{\bibfnamefont{M.}~\bibnamefont{Sferrazza}},
  \bibnamefont{and} \bibinfo{author}{\bibfnamefont{C.~E.}
  \bibnamefont{Williams}}, \bibinfo{journal}{Europhys. Lett.}
  \textbf{\bibinfo{volume}{62}}, \bibinfo{pages}{110}
  (\bibinfo{year}{2003}{\natexlab{b}}).

\bibitem[{\citenamefont{Th\'{e}odoly et~al.}(2001)\citenamefont{Th\'{e}odoly,
  Ober, and Williams}}]{Theodoly_EPJE}
\bibinfo{author}{\bibfnamefont{O.}~\bibnamefont{Th\'{e}odoly}},
  \bibinfo{author}{\bibfnamefont{R.}~\bibnamefont{Ober}}, \bibnamefont{and}
  \bibinfo{author}{\bibfnamefont{C.~E.} \bibnamefont{Williams}},
  \bibinfo{journal}{Eur. Phys. J. E.} \textbf{\bibinfo{volume}{5}},
  \bibinfo{pages}{51} (\bibinfo{year}{2001}).

\bibitem[{\citenamefont{Baigl et~al.}(2002)\citenamefont{Baigl, Seery, and
  Williams}}]{Baigl_2002}
\bibinfo{author}{\bibfnamefont{D.}~\bibnamefont{Baigl}},
  \bibinfo{author}{\bibfnamefont{T.~A.~P.} \bibnamefont{Seery}},
  \bibnamefont{and} \bibinfo{author}{\bibfnamefont{C.~E.}
  \bibnamefont{Williams}}, \bibinfo{journal}{Macromolecules}
  \textbf{\bibinfo{volume}{35}}, \bibinfo{pages}{2318} (\bibinfo{year}{2002}).

\bibitem[{\citenamefont{Essafi et~al.}()\citenamefont{Essafi, Baigl, and
  Williams}}]{Baigl_essafi}
\bibinfo{author}{\bibfnamefont{W.}~\bibnamefont{Essafi}},
  \bibinfo{author}{\bibfnamefont{D.}~\bibnamefont{Baigl}}, \bibnamefont{and}
  \bibinfo{author}{\bibfnamefont{C.~E.} \bibnamefont{Williams}},
  \bibinfo{note}{in preparation}.

\bibitem[{\citenamefont{McCormick and Chen}(1982)}]{McCormck_1982}
\bibinfo{author}{\bibfnamefont{C.~L.} \bibnamefont{McCormick}}
  \bibnamefont{and} \bibinfo{author}{\bibfnamefont{G.~S.} \bibnamefont{Chen}},
  \bibinfo{journal}{J. Polym. Sci.: Polym. Chem. Ed.}
  \textbf{\bibinfo{volume}{20}}, \bibinfo{pages}{817} (\bibinfo{year}{1982}).

\bibitem[{\citenamefont{Pershan and Als-Nielsen}(1984)}]{Pershan_1984}
\bibinfo{author}{\bibfnamefont{P.~S.} \bibnamefont{Pershan}} \bibnamefont{and}
  \bibinfo{author}{\bibfnamefont{J.}~\bibnamefont{Als-Nielsen}},
  \bibinfo{journal}{Phys. Rev. Lett.} \textbf{\bibinfo{volume}{52}},
  \bibinfo{pages}{759} (\bibinfo{year}{1984}).

\bibitem[{not({\natexlab{a}})}]{note_absorption}
\bibinfo{note}{If the same experiment had been made at 8 keV, the energy of the
  cupper K$\alpha_1$ radiation provided by a rotating anode, the absorption
  coefficient would have been $\alpha \approx 10^{-11}$.}

\bibitem[{not({\natexlab{b}})}]{note_energie}
\bibinfo{note}{Preliminary measurements were also performed at a photon energy
  of 18 to 20 keV. See \emph{ESRF highlights} 1997/1998, pp. 20-21}.

\bibitem[{not({\natexlab{c}})}]{note_AMAMPS}
\bibinfo{note}{Corresponding to these curves, the fits have provided the
  following parameters for the adsorbed AMAMPS (hydrophilic polyelectrolyte)
  monolayer: $h_{AMAMPS}\approx$ 2~nm, $\rho_{AMAMPS}\approx$
  0.35~e$^-\AA^{-3}$, $\sigma_{AMAMPS}\approx$ 1.0~nm. These parameters are
  independent of $f$, as illustrated by figure~\ref{Deltaq4I_AMAMPS}}.

\bibitem[{\citenamefont{Parratt}(1954)}]{Parat_1954}
\bibinfo{author}{\bibfnamefont{L.~G.} \bibnamefont{Parratt}},
  \bibinfo{journal}{Phys. Rev.} \textbf{\bibinfo{volume}{95}},
  \bibinfo{pages}{359} (\bibinfo{year}{1954}).

\bibitem[{\citenamefont{Nevot and Croce}(1980)}]{Nevot_1980}
\bibinfo{author}{\bibfnamefont{L.}~\bibnamefont{Nevot}} \bibnamefont{and}
  \bibinfo{author}{\bibfnamefont{P.}~\bibnamefont{Croce}},
  \bibinfo{journal}{Rev. Phys. Appl.} \textbf{\bibinfo{volume}{15}},
  \bibinfo{pages}{761} (\bibinfo{year}{1980}).

\bibitem[{not({\natexlab{d}})}]{note_density}
\bibinfo{note}{This anomalous increase was also observed in recent experiments
  on other interfaces in water ~\cite{Doer_2000,Schwendel}}.

\bibitem[{not({\natexlab{e}})}]{note_chainlength}
\bibinfo{note}{In the case of sufficiently short chains, a globular state is
  expected and \emph{h} depends on \emph{N}, as shown
  in~\cite{Baigl_pearlsize}}.

\bibitem[{\citenamefont{Dobrynin and Rubinstein}(2002)}]{Dobrynin_2002}
\bibinfo{author}{\bibfnamefont{A.~V.} \bibnamefont{Dobrynin}} \bibnamefont{and}
  \bibinfo{author}{\bibfnamefont{M.}~\bibnamefont{Rubinstein}},
  \bibinfo{journal}{Macromolecules} \textbf{\bibinfo{volume}{35}},
  \bibinfo{pages}{2754} (\bibinfo{year}{2002}).

\bibitem[{not({\natexlab{f}})}]{note_perles_sel}
\bibinfo{note}{In solution, the pearl characteristics (conformation and size)
  are insensitive to the presence of added salt, as shown by Zero Average
  Contrast Neutron Scattering (ZAC SANS) experiments. M.-N. Spitteri, PhD
  thesis, Paris XI, Orsay, France (1997)}.

\bibitem[{\citenamefont{Rayleigh}(1882)}]{Rayleigh_1882}
\bibinfo{author}{\bibfnamefont{J.~W.~S.} \bibnamefont{Rayleigh}},
  \bibinfo{journal}{Phil. Mag.} \textbf{\bibinfo{volume}{14}},
  \bibinfo{pages}{184} (\bibinfo{year}{1882}).

\bibitem[{\citenamefont{Borisov et~al.}(2001)\citenamefont{Borisov, Hakem,
  Vilgis, Joanny, and Johner}}]{Borisov_2001}
\bibinfo{author}{\bibfnamefont{O.~V.} \bibnamefont{Borisov}},
  \bibinfo{author}{\bibfnamefont{F.}~\bibnamefont{Hakem}},
  \bibinfo{author}{\bibfnamefont{T.~A.} \bibnamefont{Vilgis}},
  \bibinfo{author}{\bibfnamefont{J.-F.} \bibnamefont{Joanny}},
  \bibnamefont{and} \bibinfo{author}{\bibfnamefont{A.}~\bibnamefont{Johner}},
  \bibinfo{journal}{Eur. Phys. J. E.} \textbf{\bibinfo{volume}{6}},
  \bibinfo{pages}{37} (\bibinfo{year}{2001}).

\bibitem[{\citenamefont{Th{\'e}odoly}(1999)}]{Theo_thesis}
\bibinfo{author}{\bibfnamefont{O.}~\bibnamefont{Th{\'e}odoly}}, Ph.D. thesis,
  \bibinfo{school}{Paris VI}, \bibinfo{address}{Paris, France}
  (\bibinfo{year}{1999}).

\bibitem[{\citenamefont{Doer et~al.}(2000)\citenamefont{Doer, Tolan, Schlomka,
  and Press}}]{Doer_2000}
\bibinfo{author}{\bibfnamefont{A.~K.} \bibnamefont{Doer}},
  \bibinfo{author}{\bibfnamefont{M.}~\bibnamefont{Tolan}},
  \bibinfo{author}{\bibfnamefont{J.-P.} \bibnamefont{Schlomka}},
  \bibnamefont{and} \bibinfo{author}{\bibfnamefont{W.}~\bibnamefont{Press}},
  \bibinfo{journal}{Europhys. Lett.} \textbf{\bibinfo{volume}{52}},
  \bibinfo{pages}{330} (\bibinfo{year}{2000}).

\bibitem[{\citenamefont{Schwended et~al.}(2003)\citenamefont{Schwended,
  Hayashi, Steitz, Dahint, Pipper, Pertsin, and Grunze}}]{Schwendel}
\bibinfo{author}{\bibfnamefont{D.}~\bibnamefont{Schwended}},
  \bibinfo{author}{\bibfnamefont{T.}~\bibnamefont{Hayashi}},
  \bibinfo{author}{\bibfnamefont{R.}~\bibnamefont{Steitz}},
  \bibinfo{author}{\bibfnamefont{R.}~\bibnamefont{Dahint}},
  \bibinfo{author}{\bibfnamefont{J.}~\bibnamefont{Pipper}},
  \bibinfo{author}{\bibfnamefont{A.}~\bibnamefont{Pertsin}}, \bibnamefont{and}
  \bibinfo{author}{\bibfnamefont{M.}~\bibnamefont{Grunze}},
  \bibinfo{journal}{Langmuir} \textbf{\bibinfo{volume}{19}},
  \bibinfo{pages}{2284} (\bibinfo{year}{2003}).

\end{thebibliography}

\end{document}